\documentclass[useAMS,usegraphicx]{mn2e}

\title[X-ray eclipse of the AGN in NGC 1365]{The XMM-Newton long look of 
NGC~1365: uncovering of the
obscured X-ray source
}
\author[G. Risaliti et al.]
{G. Risaliti,$^{1,2}$ 
M.~Salvati,$^2$ M.~Elvis,$^1$ G.~Fabbiano,$^1$ A.~Baldi,$^{1}$
S.~Bianchi,$^{3}$
\newauthor V.~Braito,$^{4}$ M~Guainazzi,$^{5}$ G.~Matt,$^{3}$
G.~Miniutti,$^{6}$ J.~Reeves,$^{4}$ R.~Soria,$^{7}$ A.~Zezas$^{1}$\\
$^1$ Harvard-Smithsonian Center for Astrophysics, 60 Garden St. 
Cambridge, MA 02138 USA {E-mail: grisaliti@cfa.harvard.edu}\\
$^2$ INAF - Osservatorio di Arcetri, L.go E. Fermi 5,
Firenze, Italy\\
$^3$ Dipartimento di Fisica, Universit\`a degli Studi ``Roma Tre'',
Via della Vasca Navale 84, I-00146 Roma, Italy\\
$^4$ Astrophysics Group, School of Physical and Geographical Science, Keele University, Keele, Staffordshire ST5 5BG, UK\\
$^5$ European Space Astronomy Centre of ESA, Apartado 50727, E-28080 Madrid, Spain\\
$^6$ Laboratoire Astroparticule et Cosmologie (APC), UMR 7164, 10 Rue A.
Domon et L. Duquet, 75205 Paris\\
$^7$ MSSL, University College London, Holmbury St. Mary, Dorking, Surrey, RH5 6NT, UK}

\begin{document}

\date{Released Xxxx Xxxxx XX}

\pagerange{\pageref{firstpage}--\pageref{lastpage}} \pubyear{2002}

\maketitle

\label{firstpage}

\begin{abstract}
We present an analysis of the extreme obscuration variability observed
during an {\em XMM-Newton} 5-days continuous monitoring of the
AGN in NGC 1365. The source was in a reflection-dominated state
in the first $\sim1.5$~days, then a strong increase of the 7-10~keV
emission was   observed in $\sim$10~hours, followed by a
symmetric decrease. The spectral analysis of the different
states clearly shows that this variation is due to an uncovering
of the X-ray source. From this observation we estimate
a size of the X-ray source $D_S<10^{13}$~cm, a distance
of the obscuring clouds $R\sim10^{16}$~cm and a density $n\sim10^{11}$~cm$^{-3}$.
These values suggest that the X-ray absorption/reflection originate
from the broad line region
clouds. This is also supported by the resolved width of
the iron narrow $K\alpha$ emission line, consistent with the
width of the broad $H\beta$ line.
\end{abstract}



\section{Introduction}
The structure, size and physical properties of the circumnuclear medium of AGNs is
still a matter of debate and investigation. The axisymmetric (torus-like) geometry
required by AGN unified models (e.g. Antonucci~1993) can be obtained by means of quite
different structures, from galactic dust lanes to sub-parsec scale, high density tori.
In this context, the hard X-ray observations of the nearby obscured Seyfert 
Galaxy NGC~1365 have proven to be a unique opportunity to investigate the
properties of the AGN circumnuclear medium, as well as to put tight constraints on
the size of the X-ray source.

NGC~1365 is a nearby (z=0.0055) Seyfert 1.8 Galaxy. Several 
observations performed in the hard X-rays in the past
$\sim15$ years with all the major X-ray observatories revealed 
several spectral changes from Compton-thin
to reflection dominated states, implying 
an extraordinary
absorption variability 
from $N_H\sim10^{23}$~cm$^{-2}$ to 
$N_H>10^{24}$~cm$^{-2}$
(Risaliti et al. 2005A, 2007, hereafter R05A and R07). 
The most extreme variations have been observed during a 
recent {\em Chandra} campaign  consisting of six short (15~ks)
observations performed in ten days.
The source was in a Compton-thin state during the first observations,
then ``disappeared'' (i.e. only the reflected component was visible) in
the second observation two days later, and was back in the initial state 
during the third and later observations.
The interpretation of this extreme variability is based on the presence
of Compton-thick eclipsing clouds crossing the line of sight to the central source.
We discussed this scenario in R07 where we obtained
stringent constraints on the X-ray source size ($D<10^{14}$~cm) and, under the
hypothesis of Keplerian motion of the eclipsing clouds, on the distance of the
absorber from the central black hole ($R\leq10^{16}$~cm).\\
The main limits of this analysis are the lack of continuous monitoring during the eclipse,
and the relatively low S/N of the {\em Chandra}
spectra, due to the low effective area at high energies (E$>$2~keV) and the 
short observing time.
In order to solve these problems, we obtained
a 5-day {\em XMM-Newton} 
almost-continuous look of NGC~1365.
During this observation the source was on average in a low flux, reflection-dominated
state. However, a dramatic spectral change was observed after $\sim2$~days,
when a hard transient component appeared for $\sim60$~ks.
In this paper we present the analysis of this {\em XMM-Newton} long look
of NGC~1365, and we focus on the implications of the observed absorption variability.
\section{Data reduction and analysis}
The observation was performed from May 30 to June 5, 2007.
for three complete orbits (1384-1386). The reduction and calibration were performed
with the SAS package,
following 
standard procedures as recommended by the {\em XMM-Newton} Science Operation
Center.

We included all the time intervals in the timing and spectral analysis, 
except for the final $\sim15$~ks of the
second orbit, when high background flares saturated the EPIC detectors.
The spectra and light curves were extracted from a circular region with a 25~arcsec radius.
With this choice, the background level is always lower than 5\% of the source spectrum at all times and
energy intervals.
For both the timing and spectral analysis, we always used the complete PN and MOS (merged)
set of data, which are fully consistent with each other. In the figures however we only show
the PN data, for clarity. 


In Fig.~1 we show the low-energy (2-5~keV) and high-energy (7-10~keV) light curves
obtained from the PN.
A visual analysis of these curves reveals a strong variation during the first half
of the second orbit, consisting of an increase of the hard flux, followed by a symmetric
decrease. The hard flux in the second half of the second orbit and in the third 
orbit remains constant, at a level slightly higher than in the first orbit. 
At low energies the source remains constant throughout the whole observation.
Considering the 
past observed
fast absorption changes, the most likely interpretation of this
variability is a transient decrease of the absorbing column density during the
second orbit.

In order to test this hypothesis and to perform a complete spectral
analysis of our data, we extracted the spectra from six time intervals,
corresponding to different spectral states, as shown in Fig.~1
(the fifth and sixth interval show no obvious variations, and are separated only
because they belong to two different orbits).
The spectral analysis was performed with the XSPEC~12.4 package (Arnaud et al.~1996).

Here we are only interested in the variations observed in the high-energy (E$>$2~keV) 
emission. Therefore, we do not discuss in detail the lower energy (0.5-2~keV) emission,
 which has a diffuse origin,
as shown by {\em Chandra} images (R05A).
However, this component  cannot be completely neglected 
for two reasons: 1) an analysis of the {\em Chandra} diffuse emission shows a 
faint, but not negligible, tail at energies higher than 2 keV;
 2) the fits of the high energy spectra should not include components
whose extrapolations towards low energies over-predict the observed emission.
In order to take into account these aspects, we extracted a spectrum of the
diffuse emission from the total
{\em Chandra} data, in an annulus centered on the nucleus, with external radius
equal to that used in the {\em XMM-Newton} analysis, and inner radius of two arcsec,
in order to remove the central emission. We then fitted this spectrum with
a multi-component model providing a good analytical representation
of the data. 
The analysis of the {\em XMM-Newton} spectra was performed adding to 
all models the fixed components found in the {\em Chandra} analysis, with
a 5\% maximum range in the cross-calibration. 

We first performed the analysis of 
the six
intervals separately. The spectrum of the first interval can be entirely reproduced
by a continuum reflection (PEXRAV model in XSPEC, Magdziarz \& Zdziarski~1995)
plus a set of three emission lines, at energies between 6.4 and 7.0~keV. 
The other five spectra are successfully reproduced with the same model as above, plus an
intrinsic continuum component, consisting of an absorbed power law.
The photon index, column density and flux of the intrinsic component
were left free to vary for each spectrum. The slope of the incident continuum
for the reflection component is assumed to be the same as that of the intrinsic emission. 
In all cases the parameters of the reflection continuum
are compatible with the values found in the first spectrum. 
The column density and
photon index of the intrinsic component are also
constant within the errors, with values in the ranges $\Gamma\sim2.3-2.9$ and
$N_H\sim5-10\times10^{23}$~cm$^{-2}$.
Instead, the flux shows strong
variations.  

As a second step in order to obtain a physically consistent view of the observed variability,
we performed a simultaneous analysis of the six spectra, imposing constant reflection parameters.
 The photon index and the absorption column density of the intrinsic power law were also
required to be constant, while the flux was left free to vary in all the six intervals.
The best fit parameters are shown in Tab.~1. In Fig.~2 we show the three spectra from the first, 
third and sixth interval, and the residuals for all the intervals. It is clear from the
overall goodness of fit and from the residual distribution, showing no obvious unaccounted features, 
that the adopted model
provides a good representation of the spectral variations observed in Fig.~1.


\begin{figure}
\includegraphics[width=8.5cm]{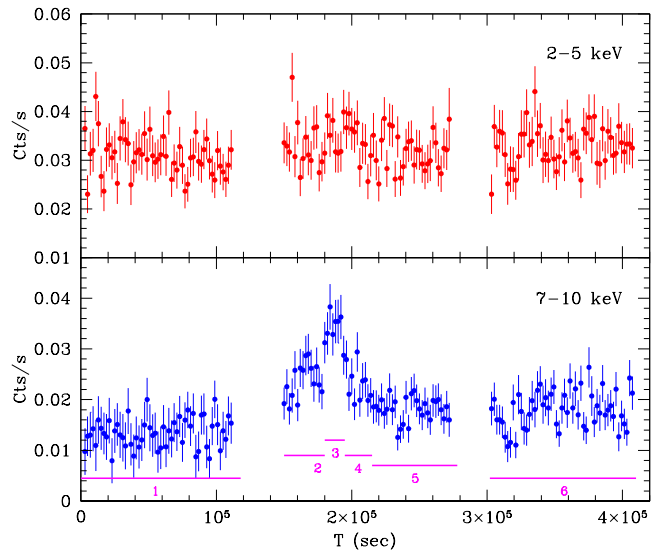}
\caption{2-5 keV (top) and 7-10 keV (bottom) light curves of the
nuclear emission of NGC 1365 during the {\em XMM-Newton} long look.
The horizontal segments show the extraction time intervals for the six
spectra discussed in the text.  
}
\end{figure}

\begin{figure}
\includegraphics[width=8.5cm]{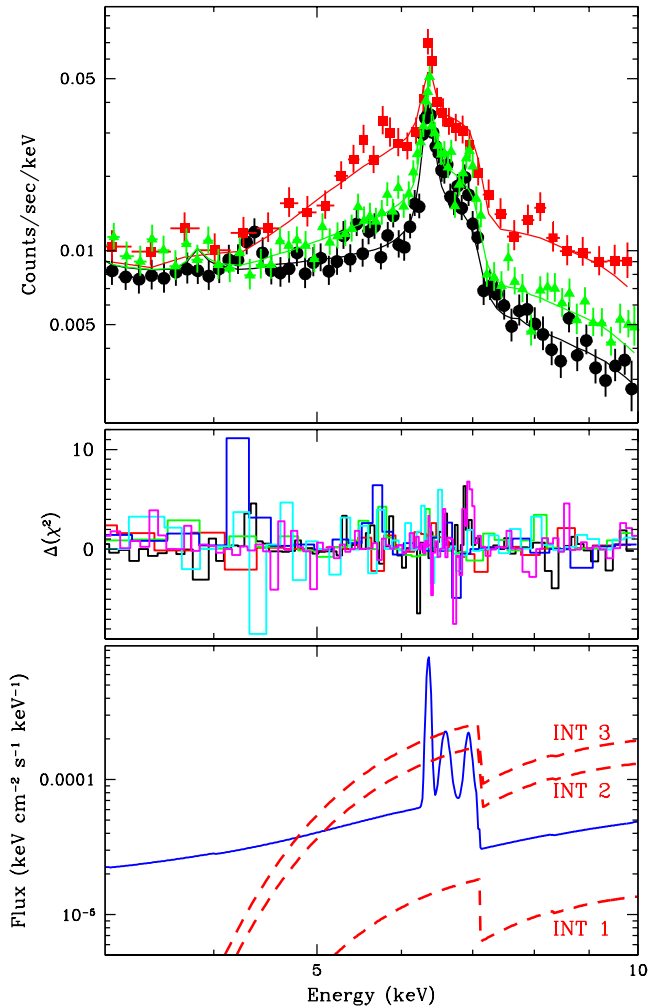}
\caption{Upper panel: PN Spectra obtained from the time intervals 1, 3 and 6 as shown in Fig.~1.
The increase in the interval~3 spectrum (red diamonds) 
and, in a smaller fraction, in the interval~6 spectrum (green triangles) is 
well reproduced by an intrinsic power law continuum absorbed by a 
column density $N_H\sim9\times10^{23}$~cm$^{-2}$ (see text for details).
The other three PN and six MOS spectra are not shown for clarity, but are fully discussed in
the text. Middle panel: contribution to $\chi^2$ for each of the six spectra, showing no
significant feature not included in the model. Lower panel: best fit model components 
for the first, second and third
interval. The continuous lines are common to all intervals, and represent the reflected
components. The dashed lines show the intrinsic component, negligible in the first interval,
showing up in the second interval, and further increasing in the third.
}
\end{figure}

\begin{table}
\caption{XMM-Newton Analysis}
\centerline{\begin{tabular}{lc|lc}
\hline
$\Gamma$ & 2.76$^{+0.37}_{-0.37}$   & Norm 1$^a$   & 0.2$^{+0.07}_{-0.07}$\\
N(Refl)$^a$  & 0.73$^{+0.4}_{-0.3}$ & Norm 2$^a$   & 1.9$^{+0.2}_{-0.2}$\\
$N_H^b$    & 94$^{+10}_{-10}$       & Norm 3$^a$   & 2.8$^{+0.3}_{-0.3}$\\
&&Norm 4$^a$   & 1.8$^{+0.2}_{-0.2}$\\
&&Norm 5$^a$   & 0.8$^{+0.1}_{-0.1}$\\
$\chi^2$/d.o.f.& 1799/1711 & Norm 6$^a$   & 0.8$^{+0.1}_{-0.1}$ \\ 
\hline
\end{tabular}}
\footnotesize{$^a$: Normalizations of the reflected component
(constant in all intervals) and of the transmitted components
in each of the six intervals shown in Fig.~1, in units of
$10^{-3}$~keV~s$^{-1}$~cm$^{-2}$~keV$^{-1}$. $^b$: Absorbing column
density of the intrinsic component, in units of 10$^{22}$~cm$^{-2}$.}
\end{table}

\begin{table}
\caption{Line Parameters}
\centerline{\begin{tabular}{lccc}
Line & E (keV) & $\sigma$ (eV) & EW$^a$ (eV)\\
\hline
1  & 6.40$^{+0.06}_{-0.05}$ & 30$^{+14}_{-19}$&800$^{+70}_{-60}$ \\
2  & 6.64$^{+0.02}_{-0.04}$ & 58$^{+34}_{-36}$&180$^{+45}_{-30}$ \\
3  & 6.95$^{+0.02}_{-0.02}$ & $<91$             &330$^{+50}_{-60}$ \\
\hline
\end{tabular}}
\footnotesize{$^a$ EWs are estimated with respect to
the reflection component.}
\end{table}
\section{Discussion}

The spectral analysis presented above shows that the observed high energy variations
are due to a power law component, absorbed by a column density 
$N_H\sim9\times10^{23}$~cm$^{-2}$, rising from
$\sim$zero up to its maximum flux in $\sim30$~ks, and then decreasing down to
$\sim20$\% of its maximum in a similar time. 
These observations can be interpreted as due to
intrinsic continuum variability, or to a change in the absorbing column density
along the line of sight.
The former case would imply a switch-on of the X-ray source from an almost completely
reflection-dominated state (Tab.~1) in $\sim30$~ks, followed
by a strong decrease in a similar time. Such short time scales are unlikely in
a disk-corona scenario, as discussed in R05A, however they cannot be completely
ruled out.
On the other side,
absorption variability is a most likely scenario in this case, given the X-ray observational
history of NGC~1365: \\
- Unequivocal column density variations in short times scales ($<2$~days)
have been observed during the {\em Chandra} campaign (R07).
Here we refer in particular to the Compton-thin variations occurred during
the last four observations, which cannot be interpreted as due to
intrinsic continuum variability.\\
- Another transit of an obscuring cloud has been detected in a previous
{\em XMM-Newton} 60~ks observation (Risaliti et al.~2008, submitted).
In this case the crossing cloud does not completely cover the source, and
its column density is $\sim3\times10^{23}$~cm$^{-2}$. This makes the
spectral analysis more complicated than in our case. However, the 
eclipsing time is of the same order as that observed here, while the underlying
continuum remains constant.\\
- Considering all the past X-ray observations where the source was in a 
Compton-thin state, strong variations of the primary continuum
have never been observed: the estimated intrinsic flux is constant within
a factor of at most two. 

Based on these arguments, we conclude that the interpretation invoking absorption variability 
is the most appropriate for our observations.
Within this scenario, 
thick clouds move along the line of sight,
uncovering and then covering again the primary X-ray source. 
The covering factor of the thick clouds is almost complete during the first {\em XMM-Newton}
orbit (INT~1 in Fig.~1), 
then decreases to a minimum, and then increases again, leaving only $\sim10$\%
of the source uncovered in INT~5 and 6. It is not possible from these observations
to tell directly whether the uncovering is complete at the flux maximum. We note however 
that the extrapolated 2-10~keV flux in INT~3 (F(2-10)$\sim3\times10^{-11}$~erg~s$^{-1}$~cm$^{-2}$, adopting
the continuum parameters shown in Table~1) is compatible with the highest
value measured in past observations (Risaliti et al.~2005B, hereafter R05B). 
This is an indication that most of the X-ray source
is probably uncovered during INT~3. 

Two schemes for the 
geometrical structure of the circumnuclear absorber are possible (Fig.~3):\\
1) The Compton-thick absorber is made of clouds with about 
the same size of the source. On average, few (1-3) of these clouds cross the
line of sight. Depending on the fluctuations in the number and column density of the clouds
the source can be seen in a reflection-dominated state for most of the time, 
with a short interval in a Compton-thin state, as observed.\\
2) The thick clouds are bigger than the X-ray source
by a factor of the order of the ratio between the continuous Compton-thick and Compton-thin intervals
in our observation. In this case our light curve can be reproduced by 
 the crossing of a single thick cloud, followed by a short interval where only thin gas
is present along the line of sight, followed by another ``big'' thick cloud.

The observations discussed here can only constrain the properties of the thick clouds,
but nothing can be directly inferred on the thin material. In principle,  
it can be in a different physical state (either homogeneous, or distributed in clouds of 
different size) and distance (either closer or farther) from the central source.
The variations of Compton-thin absorption observed with {\em Chandra} (R07) in time scales of two days
rule out a homogeneous material. Considerations on the ionization state (see below for
more details) also rule out that the thin material is closer to the center than the thick clouds.
However, a layer of thin clouds external to the thick ones, or a co-spatial mix
of thick and thin clouds, are both possible. 
\begin{figure}
\includegraphics[width=8.5cm,angle=0]{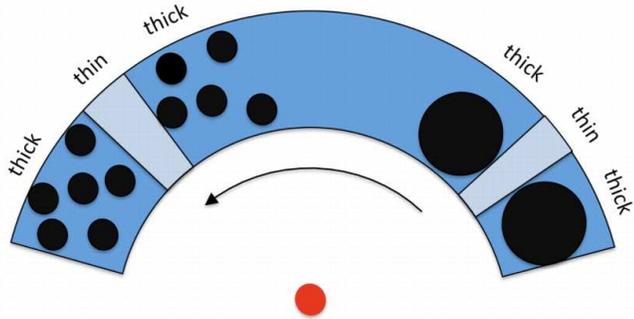}
\caption{Possible models for the structure of the circumnuclear absorber.
The black points represent Compton-thick clouds.
The regions with a darker background in the "torus" are those where at least
one thick cloud crosses the line of sight to the X-ray source
(the red spot in the center). The regions with a lighter background show
the Compton-thin lines of sight. 
}
\end{figure}

In order to estimate the relevant geometrical and 
physical quantities (i.e. the size 
of the X-ray source, the size and distance from the center of the obscuring clouds, their
density) we first note that the covering/uncovering time depends only on 
the source size and on the velocity of the thick clouds, but not on the 
dimensions of the obscuring clouds, nor on the structure of the thin gas. 
If the covering clouds (for the remainder of the
analysis, we always refer to the
Compton-thick component only) have different velocities,
the maximum size of the X-ray emitting region is still given by the velocity of
the fastest cloud, $v$, times the eclipsing time, $T$.
In R07 we discussed possible ways to constrain these two quantities:\\ 
- The time intervals between consecutive observations which caught the source in different
states provided upper limits to the eclipsing time $T<$1~day.\\
- Two main arguments provided constraints to the cloud velocity: 1) a limit to
the ionization state of the absorbing gas, which cannot be too highly ionized in order
to effectively absorb the primary component, and 2) a limit on the width of the
narrow iron emission line, under the assumption that reflection and absorption are 
due to the same physical component. 

The new observations presented here allow an improved estimate of all the 
above constraints: 1) the eclipsing time $T$ is directly measured from the
observations ($T\sim$10~hours); 2) the iron narrow emission line is now resolved 
and therefore the line width can be directly
measured; 3) having a precise determination of the reflection-dominated spectrum 
and a measurement of the intrinsic continuum from past observations
(R05B)
we can put a more stringent upper limit on the ionization parameter U of the
obscuring cloud. 
Here we discuss the line width and ionization parameter arguments, following the
scheme of R07.

{\bf 1. Ionization parameter.} In order to obtain  complete obscuration of the 
primary source, as observed in INT~1 (Fig.~1), the 
obscuring gas cannot be too highly ionized. 
The intrinsic continuum can be estimated from previous
X-ray observations. In particular, the {\em XMM-Newton} observation of January 2004,
described in R05B, provided the lowest-$N_H$, highest S/N spectrum
available. We therefore adopted the best fit continuum model from this observation
as the intrinsic ionizing emission from the central X-ray source.
We added this component to the best fit model for INT~1
and we allowed for ionized absorption, following the model by Done~(1992).
In order to reproduce the observed spectrum without over-predicting the
continuum, the ionization parameter must be $U<20$~erg~cm~s$^{-1}$, and the absorbing column 
density $N_H>2\times10^{24}$~cm$^{-2}$. 
From this limit we can obtain upper limits on the source size and on the cloud
distance from the center, $R$, under the assumption of Keplerian cloud velocity 
$v=(GM_{BH}/R)^{1/2}$, and assuming that the cloud size $D_C$
is larger than the size of the X-ray source $D_S$. 
With simple algebra (see R07 for details) 
we obtain: 
\begin{equation}
R>(GM_{BH})^{1/5}[T L_X/(U_{max}N_H)]^{2/5}
\end{equation}
where 
$L_X$ is the ionizing 1-150~keV radiation, which we take from R05B:
$L_X=1.5\times10^{42}$~erg~s$^{-1}$.
The central black hole mass, $M_{BH}$ can been estimated through the relations between
the black hole mass and (1) the bulge velocity dispersion [log$(M_{BH}/M_\odot)=7.3\pm0.4(0.3)$
Oliva et al.~1995, Ferrarese et al.~2006], and  (2) the bulge luminosity in the K-band 
[log$(M_{BH}/M_\odot)=7.8\pm0.4(0.3)$, Dong \& De Robertis~2006, Marconi \& Hunt~2003],
where the two quoted errors refer to the statistical(systematic) effects.
From these values we obtain $R>1.2\times10^{15}$~cm. The limits on
the cloud velocity and the X-ray source size are $v=(GM_{BH}/R)^{1/2}<9,000 M_7^{2/5}$~km/s,
and $D_S<2.7\times10^{13} M_7^{2/5}$~cm. 
This size can be expressed in units of gravitational radii, $R_G=2GM_{BH}/c^2$, obtaining
$D_S<10~ M_7^{-3/5} R_G$. \\
{\bf 2. Line width.} The narrow emission lines present in all spectra are due to reflection
by circumnuclear gas. It is reasonable to assume that the reflector and the absorbing clouds
are the same physical component. This is supported by the occurrence of Compton-thick states
in the past X-ray observations of NGC~1365, with a duty cycle 
of about 1/3 (Iyomoto et al.~2005, Risaliti et al.~2000, 
R05A, R07). It follows that the Compton-thick absorber covers $\sim$1/3 of
the solid angle, and therefore a significant reflection component
must be produced by this gas. Interestingly, the ratio $R$ between the 
normalizations of the reflection and intrinsic continuum components (Table~1)
in the uncovered spectrum (INT~3) is about 1/3, further supporting the identification
between the reflector and the thick absorber.
In this scenario, we can estimate the velocity of the obscuring cloud from the
width of the narrow component of the neutral iron emission line. 
Thanks to its large equivalent width in reflection dominated states, 
the long {\em XMM-Newton} observation allows a
precise analysis of the line profile, and an estimate of its width. The line is
resolved at the 4~$\sigma$ level (FWHM(Fe)=3100$_{-1900}^{+1500}$
km/s, Table 2).
The cloud velocity in the plane of the sky is then obtained
by de-projecting the 
measured line width in spherical symmetry. The final
estimate of the cloud velocity $v=2100_{-1300}^{+1000}$~km/s. The linear dimension
the X-ray source is then $D_S=6_{-4}^{+3}\times10^{12}$~cm. 
Using the black hole mass values discussed in the previous paragraph, and 
assuming Keplerian motion of the obscuring cloud, we find
$R\sim2.0\pm0.7\times10^{16}M_7^{1/5}$~cm
and $D_S=2\pm1~M_7^{-1}~R_G$. \\
In order to have a physically reasonable size of the X-ray source, $D_S\sim10~R_G$,
the black hole mass must be of the order of 2-3~$\times10^6~M_\odot$. These
values are consistent at a $\sim1~\sigma$ level with the mass estimate from
the velocity dispersion-black hole mass correlation (see above in this Section). 
The estimate from the
bulge luminosity-black hole mass relation is instead higher at a $2~\sigma$ level,
but this is expected due to the strong starburst contribution to the bulge luminosity 
of NGC~1365. A further indication of a black hole mass of the order of a few $10^6~M_\odot$
comes  
from the $M_{BH}$-$H\beta$ width correlation (Kaspi et al.~2005). 
We obtained the width of the $H\beta$ broad component from Schulz et al.~1999,
while the optical luminosity of the AGN component has been estimated
from the X-ray 2~keV intrinsic luminosity 
(R05B), assuming a standard AGN SED in the optical/UV
(Risaliti \& Elvis~2004) and the optical/X-ray ratio of Steffen et al.~(2006).
With these values $M_{BH}=2\times10^6~M_\odot$.
This estimate is rather uncertain, both on the assumptions
on the SED and on the use of the Kaspi et al. correlation down to a luminosity
lower than those for which the correlation has been tested.

The above analysis, mainly focused on the size of
the X-ray source,
also suggests an identification between the
thick circumnuclear reflector/absorber and the broad line region.
First, we note that the resolved velocity of the iron emission line is
is consistent with the width of the broad $H\beta$ emission line 
(FWHM($H\beta$)$\sim1900$~km~s$^{-1}$, Schulz et al.~1999),
analogously to what recently found
in another Seyfert Galaxy, NGC~7213, by means of 
{\em Chandra} grating spectroscopy (Bianchi et al.~2008).
Further support to this indentification comes from the estimate of the size $D_C$ and density 
$n$ of the obscuring
clouds. As shown by the configurations in Fig.~3, $D_C$ cannot be uniquely measured
from our analysis.
However, a cloud size much higher than the
source size is unlikely from a geometrical point of view.
This is also suggested by the relatively short duration ($<2$~days) 
of the eclipse observed with {\em Chandra} (R07). Assuming $D_C$=1-3$\times D_S$,
and a column density $N_H\sim2\times10^{24}$~cm$^{-2}$ we
obtain $n\sim$1-3$\times10^{11}$~cm$^{-3}$, in agreement with the estimates for 
the broad line clouds.
\section{Conclusions}
Our analysis clearly demonstrates two main points:\\
1) The size of the X-ray source is  
$D_S<10^{13}$~cm, corresponding to a few gravitational radii.\\
2) The obscuring clouds have a distance from the center and a density typical
of the broad line region. The measured iron emission
line width is consistent with the width of the broad $H\beta$ line.
This further supports
an identification between the X-ray reflector/absorber and the broad line clouds.

Our estimates rely on a simplified scheme, with
homogeneous spherical clouds moving with Keplerian velocity  around the central source.
Possible effects not considered in our analysis are a non-Keplerian component of
the cloud velocity (for example, due to an outflow),
non-spherical geometries, and 
possible $N_H$ variations within the obscuring cloud.
All these considerations make the actual numbers uncertain within a small factor, but
do not change the two general conclusions listed above.

%

\section*{Acknowledgements}
We are grateful to the anonymous referee for his/her helpful comments.
This work has been funded by NASA Grant NNX07AR90G. 


\end{document}